# LQD-RKHS-based distribution-to-distribution regression methodology for restoring the probability distributions of missing SHM data

Zhicheng Chen,[1,2,3, #]   Yuequan Bao,[1,2,3]   Hui Li[1,2,3*] and Billie F. Spencer Jr.[4]

[1]*Key Lab of Intelligent Disaster Mitigation of the Ministry of Industry and Information Technology, Harbin Institute of Technology, Harbin, China*

[2]*Key Lab of Structures Dynamic Behaviour and Control of the Ministry of Education, Harbin Institute of Technology, Harbin, China*

[3]*School of Civil Engineering, Harbin Institute of Technology, Harbin, China*

[4]*Department of Civil and Environmental Engineering, University of Illinois at Urbana-Champaign, Urbana, IL, USA*

\# *On leave at the Department of Civil and Environmental Engineering, University of Illinois at Urbana - Champaign, Urbana, IL, USA*

**ABSTRACT**

Data loss is a critical problem in structural health monitoring (SHM). Probability distributions play a highly important role in many applications. Improving the quality of distribution estimations made using incomplete samples is highly important. Missing samples can be compensated for by applying conventional missing data restoration methods; however, ensuring that restored samples roughly follow underlying distributions of true missing data remains a challenge. Another strategy involves directly restoring the probability density function (PDF) for a sensor when samples are missing by leveraging distribution information from another sensor with complete data using distribution regression techniques; existing methods include the conventional distribution-to-distribution regression (DDR) and distribution-to-warping function regression (DWR) methods. Due to constraints on PDFs and warping functions, the regression functions of both methods are estimated from the Nadaraya-Watson kernel estimator with relatively low degrees of precision. This article proposes a new indirect distribution-to-distribution regression method in the context of functional data analysis for restoring distributions of missing SHM data. PDFs are transformed to ordinary functions residing in a Hilbert space via the newly proposed log-quantile-density (LQD) transformation; the regression for distributions is realized in the transformed space via a functional regression model constructed based on the theory of Reproducing Kernel Hilbert Space (RKHS), corresponding result is subsequently mapped back to the density space through the inverse LQD transformation. Test results using field monitoring data indicate that the new method significantly outperforms conventional methods in general cases; however, in extrapolation cases, the new method is inferior to the distribution-to-warping function regression method.

1. **Introduction**

As a strong tool for managing and maintaining civil infrastructures, structural health monitoring (SHM) systems are widely applied to infrastructures for the real time monitoring of structural responses and service conditions [1-4]. However, for many reasons, data losses and corruption are a highly common phenomenon when applying SHM [5-10]. Data losses have significant influences on SHM in terms of structural health diagnoses, decision-making as well as data processing and mining. For example, Nagayama et al. [11] indicated that data loss level of 0.38% has similar effects on power spectral density as observation noise of 5%. Some condition assessment approaches are performed based on the correlation of data from two or multiple sensors [12,13], data losses can destroy such crucial information. Being informationally incomplete, missing datasets may also yield a biased inference or decision as well as cause many unpredictable pitfalls in processing and mining SHM data. Unfortunately, data losses in a structural health monitoring system are inevitable at current technology. Therefore, restoring missing data or corresponding missing information is a very crucial topic in the field of structural health monitoring; if addressed properly, the quality and reliability of subsequent automatic diagnoses and decisions can be significantly improved.

Probability distributions, as a fundamental concept of statistics and probability theory, are widely used in engineering. In-service structural systems involve various forms of uncertainty; probability distributions are a very important and useful tool for evaluating structural performance and reliability and for assisting decision-making [14-17]. In most practical cases, distributions must be estimated from observed samples and especially in regard to data-driven approaches. A distribution model directly estimated from samples with considerable data lost or noise can introduce inaccurate, misleading and distorted information, creating several potential risks for further distribution-based applications regarding structural performance, reliability and decision-making. Hence, improving the quality of distribution estimations made using incomplete samples is necessary.

On the other hand, preserving distribution information during missing data compensation is a natural requirement for many applications. For instance, the extreme value analysis (EVA) of monitored structural responses or loading effects is central to structural safety evaluations [18]. EVA is challenging to be applied due to limited samples available on extreme events; a considerable loss of monitoring data will further reduce samples with large values, and missing data compensation can ameliorate this issue. However, it should be noted that EVA is designed to determine how likely it is for an extreme event to happen, and compensated data are expected to follow the underlying distribution of true missing data, as otherwise statistical characteristics of extreme values can be altered significantly. In such cases, compensating the distribution information of missing data takes priority.

Additionally, missing data are inherently uncertain; probability distributions are some of the best tools for characterizing uncertainty. When probability distributions of missing data can be restored, substitutive samples can be repeatedly generated from a distribution to apply multiple imputation; then, parameters of corresponding statistical models can be estimated multiple times, and probability distributions or uncertainties of model parameters can then be further investigated. Hence, restoring the distribution of missing data is also meaningful in terms of better considering uncertainty.

Numerous researchers in past decades have conducted considerable work in recovering or restoring missing SHM data. Commonly tools used include compressive sampling [19-21] or $\ell_1$-minimization [5], linear regression [6], artificial neural networks (ANNs) [22,23], support vector machines (SVMs) [24], copula modelling [7], Bayesian dynamic linear models (BDLMs) [9], etc. As an emerging approach to sparse signal reconstruction, compressive sampling- or $\ell_1$-minimization-based approaches are mainly applied to recover vibration signals due to sparsity requirements. When using a linear regression-based approach, missing data are generally replaced with fitted values of the regression model; there is no guarantee that replaced data can follow a similar distribution as true missing data. Similar problems are encountered when using ANN- and SVM-based approaches due to a lack of consideration of distributions during missing data restoration. The copula-based approach is superior in utilizing correlation and distribution information to address missing data; however, such an approach itself must be complemented with the use of appropriate distribution restoration techniques given its dependence on a reliable distribution model of missing data [7]. In the BDLMs, all observations are assumed to be normally distributed. Indeed, missing distribution information can be compensated for by restoring missing samples; however, conventional means of restoring missing SHM data are limited when employed to compensate for distribution information. For example, signal reconstruction-based methods are normally deterministic and not specifically designed to recover lost stochastic responses; for methods used to manage stochastic responses, supporting means of preserving distribution information during missing data restoration are lacking.

An alternative strategy developed recently involves directly restoring the probability density function (PDF) from missing sensor samples by leveraging distribution information from another correlated sensor with complete data via distribution regression [7, 8]. Chen et al. [7] used the conventional distribution-to-distribution (DDR) method [25] to restore missing distributions for the copula-based imputation method. The authors later found that the DDR method is limited in terms of extrapolation, and thus they proposed a distribution-to-warping function regression (DWR) method for improving extrapolation performance [8]. Response variables of regression models of the DDR and DWR methods are the PDF and warping function, respectively. It should be noted that both the PDF and warping function are special functions of a function space only closed with convex combinations; thus, regression functions of the DDR and DWR are estimated from the Nadaraya-Watson kernel estimator (classical kernel regression) with the constraint of convex combinations automatically satisfied. The precision of such approaches is largely restricted by the regression function of the Nadaraya-Watson kernel estimator being a local linear smoother.

Motivated by improving precision levels, this article proposes a new indirect distribution-to-distribution regression approach for restoring distributions of missing SHM data. The newly developed log-quantile-density (LQD) transformation [26] method is utilized to transform PDFs into a Hilbert space. After transformation, PDFs are equivalently represented by ordinary functions (free from the constraints of non-negative and unit integrals). The representation function of the target PDF is first restored by a functional regression model constructed in a Reproducing Kernel Hilbert Space (RKHS); then, the inverse LQD transformation is utilized to transform the restored representation function to obtain a substitute PDF for the missing distribution. A related study of distribution prediction in Hilbert space can be found in [27], for which distributions are embedded into a Hilbert space using the kernel mean embedding technique; however, as opposed to that of LQD transformation, kernel mean embedding is irreversible. Consequently, the prediction of a target distribution described in [27] is based on a sequence of random samples generated from a greedy optimization procedure, compromising efficiency levels. In addition to the kernel mean embedding-based approach, the distribution-to-warping function regression method [8] is also less efficient because warping functions used as training samples must be separately obtained from a function space using an optimization algorithm. By contrast, with the approach presented in this article, all mathematical methodologies involved have analytical solutions. Measures are also taken to limit errors generated through numerical integration during inverse LQD transformation and to improve the scalability of the RKHS-based functional regression model.

## 2. Introduction to log-quantile-density transformation

The log-quantile-density (LQD) transformation approach proposed in [26] is a focus of this study.

Let $f(x)$ be the probability density function (PDF) of a continuous one-dimensional distribution finitely supported on $[0, 1]$, i.e.,

$$\begin{cases} f(x) > 0, \ x \in [0, 1] \\ f(x) = 0, \ x \notin [0,1] \end{cases} \tag{1}$$

The quantile function, denoted by $Q(t)$, of $f(x)$ is the inverse function of the cumulative distribution function (CDF), i.e., $Q = F^{-1}$ where $F(x) = \int_{-\infty}^{x} f(\tau)d\tau$. Note that the range of the CDF is $[0, 1]$, and thus the quantile function is also finitely supported on $[0, 1]$, i.e., $Q(t), t \in [0, 1]$. The quantile density function, denoted by $q(t)$, is defined as the derivative of the quantile function, i.e., $q(t) = \frac{d}{dt}Q(t), t \in [0, 1]$. The above four functions satisfy the following transformational rule [26]

$$q(t) = \frac{d}{dt}Q(t) = \frac{d}{dt}F^{-1}(t) = [f(Q(t))]^{-1}, t \in [0, 1] \tag{2}$$

The LQD transformation of $f(x)$ is defined as [26]

$$\psi(t) = \log(q(t)) = -\log\{f(Q(t))\}, \ t \in [0, 1] \tag{3}$$

The transformation $\psi(t)$ is an ordinary function that does not need to satisfy the constraints of a PDF (i.e., non-negative and unit integral). Moreover, $\psi(t)$ resides in a Hilbert space $L^2[0, 1]$ where $L^2[0,1]$ is the functional space formed by all square integrable functions defined on $[0, 1]$. For a detailed discussion of the theory of Hilbert space, please refer to related texts on functional analysis. $\psi(t)$ can be mapped back to the original density space via inverse LQD transformation [26], i.e.,

$$f(x) = \theta_\psi \exp\{-\psi(F(x))\}, \quad F^{-1}(x) = \theta_\psi^{-1} \int_0^x e^{\psi(s)} ds \tag{4}$$

where $\theta_\psi = \int_0^1 e^{\psi(s)} ds$. For a detailed discussion, readers are referred to [26].

## 3. Description of problem and basic assumptions

In this section, the problem of missing distribution restoration is briefly reviewed. For a more detailed discussion, readers are referred to [8].

For monitoring data of two similar sensors installed at different locations on a structure as shown in Fig. 1, data from each sensor are divided into time segments (e.g., an hour, a day, etc.). The shapes of estimated PDFs of the segment data may be correlated between monitoring sites. When sensor B is an intermittent faulty sensor, restoring the corresponding probability distribution for a time segment with missing samples is an issue of concern. The sensor with complete data (e.g., sensor A) is used as a collaborator of the distribution restoration problem and is termed the collaborative sensor in this article.

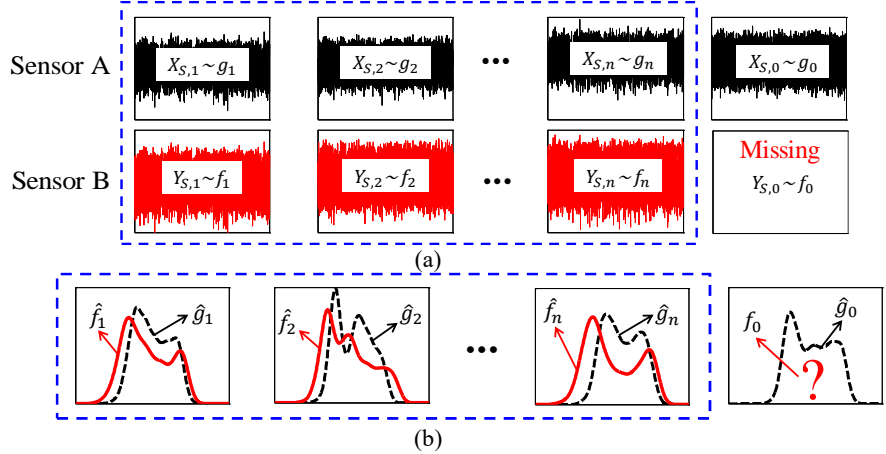

**Fig. 1.** Illustration of distribution restoration, i.e., restoring the PDF $f_0$ from missing samples by leveraging information of PDFs $\{g_i, f_i\}_{i=1}^n \cup g_0$ with complete samples [8]: (a) segments of the monitoring data and (b) estimated PDFs of corresponding segment data.

When the missing distribution can be restored, random samples generated from this distribution can be used to compensate for missing samples (e.g., missing large-value samples used for extreme value modelling). When examined data are correlated through time (i.e., structured times series), missing records can be restored by combining the distribution restoration method with the copula-based imputation method, but this involves the use of a correlated time series measured by another sensor (inter-sensor correlation) with complete records to serve as a collaborative time series. An means of restoring missing time series of strain monitoring data by combining distribution restoration and the copula-based imputation method [28]. The distribution restoration method is mainly applied to monitoring data that can be described by random variables. For acceleration data, several signal reconstruction methods (e.g., compressive sampling [19-21] or $\ell_1$-minimization [5]) have been successfully applied to recover missing records from a deterministic perspective; the distribution restoration method presented in this article is not designed for acceleration data.

Let $\{\hat{g}_i, \hat{f}_i\}_{i=1}^n \cup \hat{g}_0$ be the estimated PDFs using the corresponding segment data (see Fig. 1) as samples, similar to the previous work in [8], we will design a distribution-to-distribution regression model, respectively, using $\{\hat{f}_i, \hat{g}_i\}_{i=1}^n$ as training distributions and $\hat{g}_0$ as the predictor to obtain a prediction for $f_0$. Thus, the missing PDF can be replaced with the predicted result as its restoration.

All investigated distributions are assumed to be univariate continuous distributions finitely supported on $[0, 1]$, and distributions supported on general finite intervals can be easily managed by transforming to $[0, 1]$ via reversible scale transformation (see Appendix 4 of [8]).

The support of a distribution class may not be directly observed and must be estimated from samples. Taking sensor A shown in Fig. 1 as an example, let random variable $X_g$ denote data prior to segment division; the original support denoted by $[\omega_L^g, \omega_U^g]$ of the distribution class can then be estimated from [29-30]

$$\hat{\omega}_L^g = \min(X_g) - \kappa_L \frac{s_{X_g}}{\sqrt{n_g}} \text{ and } \hat{\omega}_U^g = \max(X_g) + \kappa_U \frac{s_{X_g}}{\sqrt{n_g}} \tag{5}$$

where $s_{X_g}$ and $n_g$ are the sample standard deviation and the sample size of $X_g$, respectively. $\kappa_L \geq 1$ and $\kappa_U \geq 1$ are two parameters controlling the impact of the sample standard deviation in estimating the support; the larger of $\kappa_L$ and $\kappa_U$, the estimated support will be wider than the interval $[\min(X_g), \max(X_g)]$. From the scale transformation presented in Appendix 4 in [8], the original support $[\hat{\omega}_L^g, \hat{\omega}_U^g]$ can be further transformed to $[0, 1]$.

Generally, the finite-support assumption of distributions can be satisfied in engineering applications, as structural responses will not tend to infinity.

## 4. Restoration method

### 4.1 Pre-processing for improving the integral accuracy of inverse LQD transformation

Generally, the integral of the inverse LQD transformation (see Eq. (4)) is frequently calculated using numerical methods. However, the numerical integration of inverse LQD transformation can generate significant errors for a PDF close to zero. To clearly understand this point, consider a beta distribution, i.e., $f(x) = B(x|\alpha_B, \beta_B)$ with parameters $\alpha_B = 6$ and $\beta_B = 3$, as shown in Fig. 2 (a). $f(x)$ of start interval $[0, x_\delta](x_\delta \approx 0.15)$ is close to zero. Eq. (2) shows that the quantile density function $q(t) = [f(Q(t))]^{-1}$ will go to infinity within interval $t \in [0, t_\delta]$ (see Fig. 2 (b)). Note that the function to be integrated in the inverse LQD transformation (i.e., $e^{\psi(t)}$) is actually the quantile density function, i.e., $e^{\psi(t)} = q(t)$ (derived from Eq. (3)). Therefore, if the quantile density function $q(t)$ of an interval goes to infinity, the integral of the inverse LQD transformation calculated via numerical methods may introduce considerable errors (see Fig. 2 (c)).

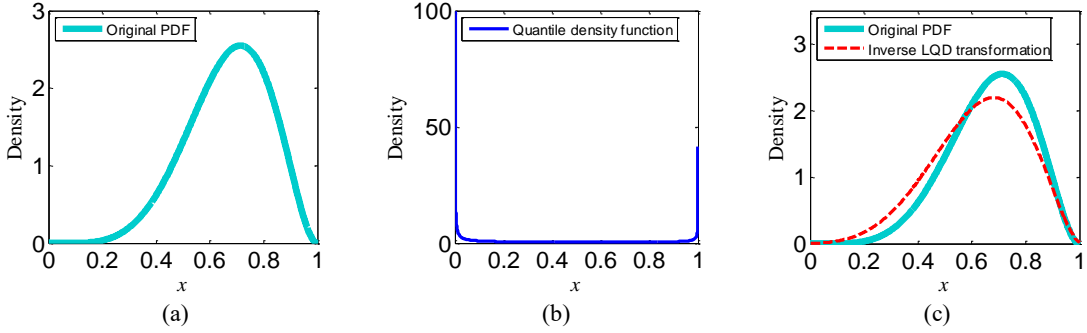

**Fig. 2.** Example showing errors generated through numerical integration during inverse LQD transformation: (a) the original PDF, (b) the quantile density function, and (c) a comparison of the original PDF (solid curve) and the PDF obtained via inverse LQD transformation through numerical integration (dashed curve).

In this study, to address this problem, the original PDF is pre-processed by adding a PDF of the uniform distribution defined on $[0, 1]$, i.e.,

$$f^*(x) = (1 - \alpha)f(x) + \alpha u(x), \ 0.2 \leq \alpha \leq 0.5, \text{ and } x \in [0, 1] \tag{6}$$

where $\alpha$ is the mixture weight, and $u(x)$ is the PDF of the uniform distribution defined on $[0, 1]$, i.e., $u(x) = 1, x \in [0, 1]$. In the distribution restoration problem, the mixture PDF $f^*(x)$ is used to replace $f(x)$. Note that $u(x) = 1, \forall x \in [0, 1]$ and $f(x) \geq 0, \forall x \in [0, 1]$, and hence $f^*(x) \geq \alpha > 0$ and the quantile density function of $f^*(x)$ will no longer tend to infinity. The mixture weight $\alpha$ controls the share of the uniform distribution in the mixture distribution, which should be restricted in $[0.2, 0.5]$. When $\alpha$ is too small, it cannot effectively remedy the near-zero problem of the original PDF; when $\alpha$ is too large, it will overly dilute features of the original PDF.

To recover the original PDF, the following method can be used to eliminate effects of the added uniform distribution. Suppose that $\hat{f}^*(x)$ is an estimate of the unknown PDF $f^*(x), x \in [0,1]$; then, the estimate of the original unknown PDF $f(x), x \in [0,1]$ can be roughly recovered from $\hat{f}^*(x)$ with

$$\hat{f}(x) = \frac{1}{W}\left|\frac{\hat{f}^*(x) - \alpha u(x)}{1 - \alpha}\right|, \ W = \int_0^1 \left|\frac{\hat{f}^*(\tau) - \alpha u(\tau)}{1 - \alpha}\right| d\tau \tag{7}$$

where $W$ is the normalization coefficient.

**4.2 LQD transformation and regression scheme**

After the above pretreatment, all PDFs, i.e., $\{\hat{g}_i, \hat{f}_i\}_{i=1}^n \cup \hat{g}_0$, are transformed to the Hilbert space via LQD transformation (see Eq. (3)) as follows

$$\psi_i^{f^*}(t) = -\log\left\{f_i^*\left(Q_{f,i}^*(t)\right)\right\}, \ i = 1,2,\cdots,n \tag{8a}$$

$$\psi_i^{g^*}(t) = -\log\{g_i^*(Q_{g,i}^*(t))\}, \ i = 0,1,\cdots,n \tag{8b}$$

where $f_i^* = (1-\alpha)\hat{f}_i + \alpha u$, $g_i^* = (1-\alpha)\hat{g}_i + \alpha u$, $Q_{f,i}^*$ and $Q_{g,i}^*$ are the quantile functions of $f_i^*$ and $g_i^*$, respectively.

Generally, there are two frameworks of nonparametric functional regression: classical kernel regression using the Nadaraya-Watson kernel estimator to estimate the regression function whereby the Nadaraya-Watson kernel estimator is subjected to linear smoothing and regression function construction in a Reproducing Kernel Hilbert Space (RKHS), which generally requires that regression objects reside in a Hilbert space. In this work, PDFs are transformed to a Hilbert space and represented by ordinary functions. Therefore, the distribution restoration method can be applied using the RKHS-based nonparametric regression technique. The RKHS-based function-to-function regression method proposed in [31] presents scalability problems. For a considerably larger dataset, the method can generate insufficient memory errors and failures, as continuous functional data are infinite or high dimensional data (when represented by values of regular grids). To overcome these drawbacks, one can apply dimension reduction to functional data of the intermittent faulty sensor, i.e., $\{\psi_i^{f^*}(t)\}_{i=1}^n$. Note that functional data of the collaborative sensor, i.e., $\{\psi_i^{g^*}(t)\}_{i=0}^n$, can remain unchanged (the reasons for this will be described below).

### 4.3 Dimension reduction by functional principal component analysis (FPCA)

Functional principal component analysis (FPCA) is an effective tool for reducing the dimensions of functional data (for a detailed discussion of FPCA, readers are referred to [32]). Therefore, FPCA is employed to reduce the dimensions of functional data drawn from the intermittent faulty sensor, i.e., $\{\psi_i^{f^*}(t)\}_{i=1}^n$, $t \in [0,1]$. The original functional data are first transformed to centred functional data by subtracting the mean function; then, FPCA can be conducted by solving the eigenanalysis problem of the covariance operator of the centred functional data.

From the investigated functional dataset $\{\psi_i^{f^*}(t)\}_{i=1}^n$ defined on $[0, 1]$, let $\{\tilde{\psi}_i^{f^*}(t)\}_{i=1}^n$ be the corresponding centred functional dataset obtained from

$$\tilde{\psi}_i^{f^*}(t) = \psi_i^{f^*}(t) - \mu_{\psi f^*}(t), \ i = 1,2,\cdots,n \tag{9}$$

where $\mu_{\psi f^*}(t)$ is the mean function of the original functional dataset, i.e. $\mu_{\psi f^*}(t) = n^{-1}\sum_{i=1}^n \psi_i^{f^*}(t)$.

The covariance function and covariance operator are two critical mathematical concepts of FPCA. The covariance function, denoted by $v(t,s)$, of the centred functional dataset is defined as [32]

$$v(t,s) = n^{-1}\sum_{i=1}^{n}\tilde{\psi}_i^{f^*}(t)\tilde{\psi}_i^{f^*}(s), \quad (t,s) \in [0,1]\times[0,1] \tag{10}$$

and the covariance operator, denoted by $V_C$, is defined as the following integral transform

$$V_C\varphi = \int v(\cdot,s)\varphi(s)ds \quad \Leftrightarrow \quad (V_C\varphi)(t) = \int v(t,s)\varphi(s)ds \tag{11}$$

Then, the FPCA can be realized by solving the following functional eigenanalysis problem

$$V_C\varphi = \rho\varphi \quad \text{subject to} \quad \|\varphi\|^2 = \int \varphi^2(\tau)d\tau = 1 \tag{12}$$

where $\rho$ and $\varphi$ are the eigenvalue and eigenfunction, respectively. Several approaches can be applied to solve the above functional eigenanalysis problem [32], and the discretization approach presented in Appendix 1 is adopted in this work. Let $\rho_1 \geq \rho_2 \geq \cdots \rho_k \geq \cdots$ be the solutions of eigenvalues in descending order. Corresponding engenfunctions then form a functional orthonormal basis of the Hilbert space in which the centred functional dataset (i.e., $\{\tilde{\psi}_i^{f^*}(t)\}_{i=1}^{n}$) resides, i.e.,

$$\{\varphi_k(t)\}_{k=1}^{\infty} \text{ satisfying } \int \varphi_j(\tau)\varphi_k(\tau)d\tau = \begin{cases} 1, & if\ j = k \\ 0, & otherwise \end{cases} \tag{13}$$

According to the Karhunen–Loève theorem [32-34], the centred function $\tilde{\psi}_i^{f^*}(t)$ takes the following form

$$\tilde{\psi}_i^{f^*}(t) = \psi_i^{f^*}(t) - \mu_{\psi^{f^*}}(t) = \sum_{k=1}^{\infty}\xi_{i,k}^{f^*}\varphi_k(t), \quad i = 1,2,\cdots,n \tag{14}$$

where $\xi_{i,k}^{f^*} = \int_0^1 \tilde{\psi}_i^{f^*}(\tau)\varphi_k(\tau)d\tau$ is the principal component of $\tilde{\psi}_i^{f^*}(t)$ corresponding to $\varphi_k(t)$.

The dimension reduction of the functional data is based on the assumption that a function can be represented by the first few eigenfunctions, which yield the following truncated Karhunen–Loève representation

$$\psi_i^{f^*}(t) = \mu_{\psi^{f^*}}(t) + \sum_{k=1}^{m}\xi_{i,k}^{f^*}\varphi_k(t), \quad i = 1,2,\cdots,n \tag{15}$$

where $m$ is the truncation order. Note that both the mean function $\mu_{\psi^{f^*}}$ and eigenfunctions $\{\varphi_k\}_{k=1}^{m}$ in Eq. (15) are the same for all functions of the investigated functional dataset (i.e., $\{\tilde{\psi}_i^{f^*}(t)\}_{i=1}^{n}$); thus, a continuous function $\psi_i^{f^*}(t)$ can be approximately represented by a m-dimensional row vector formed by coefficients of the truncated Karhunen–Loève representation, i.e., $\xi_i^{f^*} = [\xi_{i,1}^{f^*}\ \xi_{i,2}^{f^*}\ \cdots\ \xi_{i,m}^{f^*}] \in \mathrm{R}^{1\times m}$, thereby realizing the dimension reduction of the functional data. Vector $\xi_i^{f^*} \in \mathrm{R}^{1\times m}$ is used in terms of the representation vector of function $\psi_i^{f^*}$ in this article. Note that the representation vector also resides in a Hilbert space, i.e., the m-dimensional real vector space $\mathrm{R}^{1\times m}$.

### 4.4 Function-to-vector regression model

After LQD transformation and FPCA-based dimension reduction, information on available

PDFs, i.e., $\{\hat{g}_i, \hat{f}_i\}_{i=1}^{n} \cup \hat{g}_0$, is transformed into the following structured dataset

$$\left\{\begin{matrix} \xi_1^{f^*} \\ \psi_1^{g^*}(t) \end{matrix}\right\}, \left\{\begin{matrix} \xi_2^{f^*} \\ \psi_2^{g^*}(t) \end{matrix}\right\}, \cdots, \left\{\begin{matrix} \xi_n^{f^*} \\ \psi_n^{g^*}(t) \end{matrix}\right\}, \left\{\begin{matrix} \text{missing} \\ \psi_0^{g^*}(t) \end{matrix}\right\} \quad (16)$$

For missing distribution restoration, the remaining task is to design a function-to-vector regression model for the structured dataset given in Eq. (16), which takes the following form

$$\xi^{f^*} = F_{reg}(\psi^{g^*}) + \varepsilon, \quad \xi^{f^*} \in \mathrm{R}^{1 \times m} \text{ and } \psi^{g^*} \in L^2[0,1] \quad (17)$$

where $F_{reg}$ is the regression function, $\varepsilon$ is an error term, $\mathrm{R}^{1 \times m}$ is the m-dimensional real vector space, and $L^2[0,1]$ is the functional space formed by all square integrable functions defined on $[0,1]$. Both $\mathrm{R}^{1 \times m}$ and $L^2[0,1]$ are Hilbert spaces.

The regression function of Eq. (17) maps from one Hilbert space (i.e., $L^2[0,1]$) to another (i.e., $\mathrm{R}^{1 \times m}$), and the unknown regression function can be constructed in a Reproducing Kernel Hilbert Space (RKHS). A RKHS denoted as $\mathrm{H}(K_{rep})$ is a special Hilbert space generated by a reproducing kernel $K_{rep}$ and endowed with an inner product $\langle \cdot, \cdot \rangle_{\mathrm{H}(K_{rep})}$; one-to-one correspondence occurs between a reproducing kernel $K_{rep}$ and RKHS $\mathrm{H}(K_{rep})$. Further information on the theory of the RKHS or of the reproducing kernel falls beyond the scope of this article; readers are referred to [31], [35], and [36] for more information.

In the RKHS framework, the unknown regression function $F_{reg}$ is assumed to reside in a RKHS $\mathrm{H}(K_{rep})$ with reproducing kernel $K_{rep}$, and $F_{reg}$ can be estimated by solving the following penalized minimization problem

$$\min_{F_{reg} \in \mathrm{H}(K_{rep})} \sum_{i=1}^{n} \left\| \xi_i^{f^*} - F_{reg}\left(\psi_i^{g^*}\right) \right\|_2^2 + \lambda_s \|F_{reg}\|_{\mathrm{H}(K_{rep})}^2 \quad (18)$$

where $\|\cdot\|_2$ is the 2-norm of vectors (i.e., $\|\xi\|_2 = \sqrt{\xi_1^2 + \cdots + \xi_m^2}, \forall \xi \in \mathrm{R}^{1 \times m}$), $\|\cdot\|_{\mathrm{H}(K_{rep})}$ is the RKHS norm induced by the inner product $\langle \cdot, \cdot \rangle_{\mathrm{H}(K_{rep})}$ (see [31] for details). The last term of the minimization problem is a regularization term whose role is to prevent overfitting; regularized parameter $\lambda_s$ is also used in terms of the smoothing parameter, as it controls the smoothness of the regression function.

According to the representer theorem [31, 37], the solution of the above minimization problem takes the following form

$$\hat{F}_{reg}(\cdot) = \sum_{j=1}^{n} K_{rep}\left(\cdot, \psi_j^{g^*}\right) \boldsymbol{\beta}_j \Leftrightarrow \hat{F}_{reg}(\psi^{g^*}) = \sum_{j=1}^{n} K_{rep}\left(\psi^{g^*}, \psi_j^{g^*}\right) \boldsymbol{\beta}_j \quad (19)$$

where $\boldsymbol{\beta}_j$ refers to undetermined vector coefficients with the same dimensions as $\xi_j^{f^*}$, and $K_{rep}$ is a reproducing kernel corresponding to the RKHS $\mathrm{H}(K_{rep})$. For the function-to-vector regression model, the reproducing kernel is an operator-valued kernel that maps from the functional space $L^2[0,1]$ to vector space $\mathrm{R}^{1 \times m}$, i.e., $K_{rep}(\cdot, \psi_j^{g^*}): L^2[0,1] \mapsto \mathrm{R}^{1 \times m}$. A commonly used operator-valued kernel is the Gaussian operator kernel, i.e.,

$$K_{rep}\left(\psi^{g^*}, \psi_j^{g^*}\right) = \exp\left\{-\frac{1}{2\sigma^2}\int \left|\psi^{g^*}(\tau) - \psi_j^{g^*}(\tau)\right|^2 d\tau\right\} I_{\text{identity}} \qquad (20)$$

where $I_{\text{identity}}$ is the identity operator of vectors (i.e., $I_{\text{identity}}\boldsymbol{\beta} = \boldsymbol{\beta}$, $\forall \boldsymbol{\beta} \in R^{1\times m}$), and $\sigma$ is the parameter of the Gaussian operator kernel. A related discussion on the design of different types of operator-valued kernels can be found in [36].

From Eq. (19) it can be seen that only undetermined items of the regression function are $\{\boldsymbol{\beta}_j\}_{j=1}^n$ where $\boldsymbol{\beta}_j$ is a $m$-dimensional row vector of $\boldsymbol{\beta}_j = [\beta_{j,1}\ \beta_{j,2}\ \cdots\ \beta_{j,m}]$; note that $m$ is also the dimension of the response vector of the regression model, i.e., $\boldsymbol{\xi}_i^{f^*} \in R^{1\times m}$. $\{\boldsymbol{\beta}_j\}_{j=1}^n$ can be arranged as the following matrix

$$\mathbf{B} = [\boldsymbol{\beta}_1^T\ \boldsymbol{\beta}_2^T\ \cdots\ \boldsymbol{\beta}_n^T]^T \in R^{n\times m} \qquad (21)$$

Using a similar derivation procedure as that of [31] to develop the RKHS-based function-to-function regression model, the minimization problem given in Eq. (18) can be equivalently transformed into the following matrix form

$$\min_{\mathbf{B}} \left\{\text{trace}\left((\mathbf{Y} - \mathbf{AB})(\mathbf{Y} - \mathbf{AB})^T\right) + \lambda_s \text{trace}(\mathbf{ABB}^T)\right\} \qquad (22)$$

where $\mathbf{Y} = \left[\left(\boldsymbol{\xi}_1^{f^*}\right)^T\ \left(\boldsymbol{\xi}_2^{f^*}\right)^T\ \cdots\ \left(\boldsymbol{\xi}_n^{f^*}\right)^T\right]^T \in R^{n\times m}$ and $\mathbf{A}$ is a $n\times n$ matrix formed by the kernel function. For example, if the Gaussian operator kernel defined in Eq. (20) is adopted, then matrix $\mathbf{A}$ is

$$\mathbf{A} = \{a_{ij}\}_{i=1\ j=1}^{n\quad n}, \text{ where } a_{ij} = \exp\left\{-\frac{1}{2\sigma^2}\int \left|\psi_i^{g^*}(\tau) - \psi_j^{g^*}(\tau)\right|^2 d\tau\right\} \qquad (23)$$

The time complexity of calculating matrix $\mathbf{A}$ from training functional data $\{\psi_i^{g^*}(t)\}_{i=1}^n$ is $O(n^2)$.

The analytical solution of $\mathbf{B}$ of the minimization problem of Eq. (22) is

$$\text{vec}(\mathbf{B}) = [(\mathbf{I}_{m\times m} \otimes \mathbf{A}) + \lambda_s \mathbf{I}_{mn\times mn}]^{-1} \text{vec}(\mathbf{Y}) \qquad (24)$$

where $\mathbf{I}_{m\times m}$ and $\mathbf{I}_{mn\times mn}$ are identity matrixes, $\otimes$ is the Kronecker product of the matrixes, and $\text{vec}(\cdot)$ is the vectorization operation of the matrixes, i.e.,

$$\text{vec}(\mathbf{U}) = (u_{11}\ u_{21}\ \cdots\ u_{n1}\ u_{12}\ u_{22}\ \cdots\ u_{n2}\ \cdots\ u_{1m}\ u_{2m}\ \cdots\ u_{nm})^T, \forall \mathbf{U} \in R^{n\times m} \qquad (25)$$

The main storage bottleneck experienced when solving the RKHS-based regression model concerns storing matrix $(\mathbf{I}_{m\times m} \otimes \mathbf{A})$. It can be observed that the dimension of matrix $(\mathbf{I}_{m\times m} \otimes \mathbf{A})$ is $mn \times mn$ where $m$ is the dimension of the representation vector of $\psi_i^{f^*}(t)$ (i.e., $\boldsymbol{\xi}_i^{f^*} = [\xi_{i,1}^{f^*}\ \xi_{i,2}^{f^*}\ \cdots\ \xi_{i,m}^{f^*}]$) and $n$ is the number of training functional samples (i.e., $\{\psi_i^{g^*}(t)\}_{i=1}^n$). The treatment of dimension reduction of $\{\psi_i^{f^*}(t)\}_{i=1}^n$ drawn from the FPCA technique (see Eq. (15)) is very meaningful. When $\psi_i^{f^*}(t)$ is represented by values of a regular grid $\{t_1, t_2, \cdots, t_m\}$ on $[0, 1]$, dense grids are needed to characterize a complex continuous function in consideration of the integral calculation of the inverse LQD transformation; then, a considerable amount of memory is lost in storing matrix $(\mathbf{I}_{m\times m} \otimes \mathbf{A})$ with dimension $mn \times mn$. When numerous training functional samples are involved, an approach not involving dimension reduction will easily spur insufficient memory errors and failures.

The dimension of matrix $\mathbf{A}$ calculated from Eq. (23) is $n \times n$ where $n$ is the number of training functions of the collaborative sensor, i.e., $\left\{\psi_i^{g^*}(t)\right\}_{i=1}^n$. The dimension reduction of the functional data $\psi_i^{g^*}(t)$ cannot reduce the size of matrix $\mathbf{A}$ and thus does not address storage bottlenecks experienced in storing matrix $(\mathbf{I}_{m\times m} \otimes \mathbf{A})$ when solving the RKHS-based regression model using Eq. (24). Additionally, dimension reduction will result in information loss and in additional computation requirements. For these reasons, functional data $\left\{\psi_i^{g^*}(t)\right\}_{i=1}^n$ (used as predictors in the regression model) are directly used in the regression model with their complete information preserved.

### 4.5 Missing distribution restoration

With the above function-to-vector regression model, the missing representation vector $\boldsymbol{\xi}_0^{f^*}$ corresponding to the missing PDF $f_0^*$ can be restored from

$$\hat{\boldsymbol{\xi}}_0^{f^*} = \hat{F}_{reg}\left(\psi_0^{g^*}\right) = \sum_{j=1}^n K_{rep}\left(\psi_0^{g^*}, \psi_j^{g^*}\right)\boldsymbol{\beta}_j \tag{26}$$

Thus, the representation function of the missing PDF $f_0^*$ of the Hilbert space can be reconstructed from the truncated Karhunen–Loève representation of Eq. (15) as follows

$$\hat{\psi}_0^{f^*}(t) \approx \mu_{\psi^{f^*}}(t) + \sum_{k=1}^m \hat{\xi}_{0,k}^{f^*}\varphi_k(t) \tag{27}$$

The missing PDF $f_0^*(x)$ can then be restored by applying the inverse LQD transformation to $\hat{\psi}_0^{f^*}(t)$ using Eq. (4). The original missing PDF $f_0(x)$ can then be restored by eliminating the effect of the added uniform distribution using Eq. (7). This distribution restoration method is based on the LQD transformation and the RKHS-based nonparametric functional regression, essentially belonging to indirect distribution-to-distribution regression approaches; thus, it is termed the LQD-RKHS method in this article.

## 5. Validation and performance evaluation

### 5.1 Monitoring data

In this section, strain monitoring data from a long-span cable-stayed bridge in China are used to verify the effectiveness and performance of the proposed LQD-RKHS method. The investigated dataset is the same as that used in a previous work [8]. Only a brief introduction to the examined data is given in this section (for more detailed information, please refer to [8]).

As is shown in Fig. 3, the monitoring strain of two longitudinal strain gauges welded onto the bottom plate of a steel girder are used in this study.

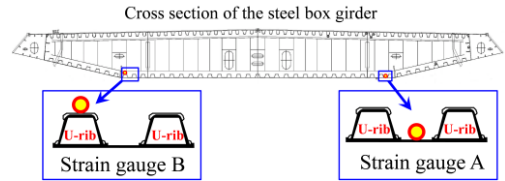

**Fig. 3.** Layout of the two examined strain gauges [8].

A total of 178 days of monitoring data (March to August of 2012) were considered. As is shown in Fig. 4, the original data were pre-processed by removing seasonal trends (see [8] for our reasons of removing seasonal trends) and then mapping to [0, 1]. Seasonal trends were estimated using the LOESS method, which is a local smoothing technique that uses weighted linear least squares and a 2nd degree polynomial model. The algorithm of the LOESS method was applied through MATLAB's "smooth" function; by tuning the span parameter, trends of different scales (i.e., seasonal, weekly, daily, etc.) reflected in raw data can be conveniently estimated.

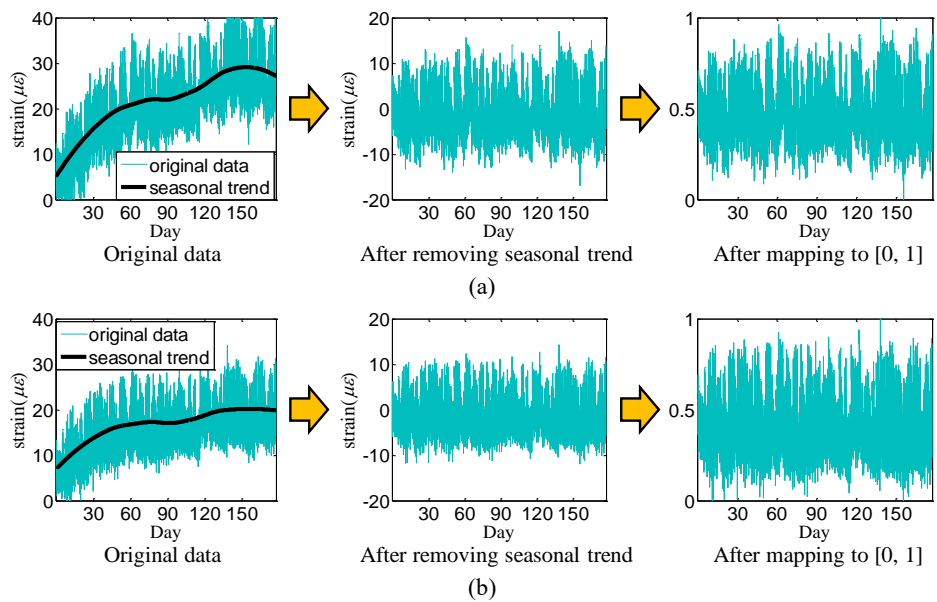

**Fig. 4.** Investigated strain monitoring data [8]: (a) monitoring data collected by Strain gauge A and (b) monitoring data collected by Strain gauge B.

Although seasonal trends have been removed, effects of daily temperature changes (i.e., temperature variations observed in the day and at night) remain in the data and are reflected as trends varying over the daily cycle. Based on slow changes in temperature, monitoring data for each hour collected by the same sensor can be assumed to follow the same distribution; however, PDFs estimated from hourly measurements are simple-shaped PDFs. For distribution restoration tests designed for effectiveness verification and performance evaluation, PDFs of complex shapes are preferable. Note that the probability distribution estimated from daily measurement data can be

**13 / 25**

regarded as a mixture distribution formed by distributions corresponding to hourly segment data, i.e.,

$$f_{\text{day}}(x) = \sum_{k=1}^{24} w_k f_{\text{hour},k}(x), \text{ where } w_k \geq 0 \text{ and } \sum_{k=1}^{24} w_k = 1 \tag{28}$$

Shapes of such mixture PDFs estimated from daily segment data are much more complex; thus, time series of the monitoring strain were divided into daily segment data (from 0:00 to 23:59 every day). Strain gauges B and A were assumed to be the intermittent faulty sensor and collaborative sensor, respectively.

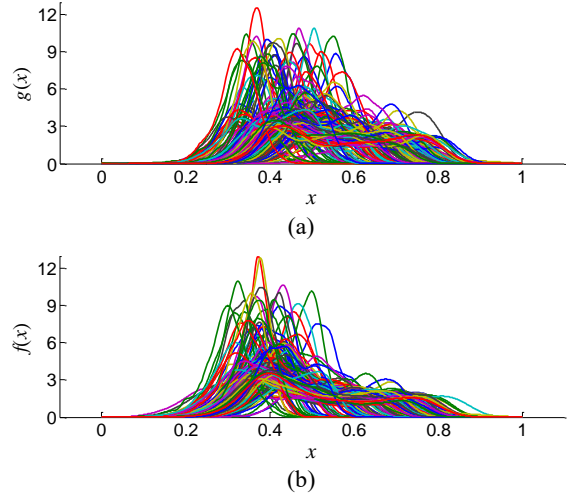

**Fig. 5.** One hundred seventy-eight estimated daily PDFs [8]: (a) PDFs of monitoring data drawn from Strain gauge A and (b) PDFs of monitoring data drawn from Strain gauge B.

The PDFs were estimated using the kernel density estimation "density" function available through the R software programme (https://cran.r-project.org/), and default settings were adopted (i.e., using Gaussian kernels with a smoothing bandwidth determined by Silverman's Rule of Thumb [38]). As originally estimated PDFs may fail to satisfy the constraint finitely supported on [0, 1], they were further truncated to [0, 1] with the following

$$\hat{f}(x) = \frac{1}{\int_0^1 \hat{f}_{\text{KDE}}(\tau) d\tau} \hat{f}_{\text{KDE}}(x) \text{I}\{0 \leq x \leq 1\} \tag{29}$$

where I{·} is the indicator function and $\hat{f}_{\text{KDE}}(x)$ is the original kernel density estimate. The 178 estimated daily PDFs (after truncation) of monitoring data drawn from Strain gauge A and Strain gauge B are shown in Fig. 5.

**5.2 Validation**

Fifty PDF pairs were randomly selected from the 178 daily PDF pairs as training PDFs, and 40 PDF pairs were randomly selected from the remaining 128 PDF pairs as test PDFs. $\{\hat{g}_i, \hat{f}_i\}_{i=1}^{50}$ and $\{\hat{g}_{0,v}, \hat{f}_{0,v}\}_{v=1}^{40}$ denote functional datasets of training PDFs and test PDFs, respectively, where $\{\hat{f}_{0,v}\}_{v=1}^{40}$ were assumed to be missing PDFs.

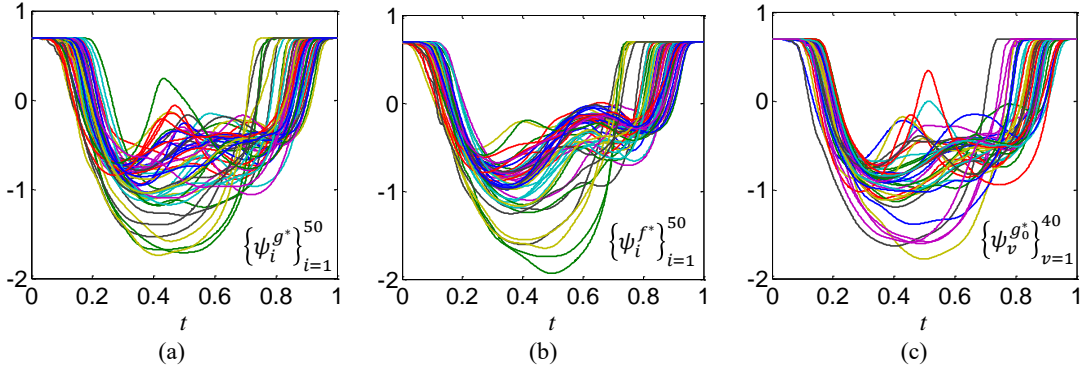

**Fig. 6.** Results transformed by the LQD transformation.

To prevent the quantile density function from tending to infinity, all investigated PDFs except for missing PDFs were pre-processed using Eq. (6), and the mixture weight $\alpha$ was set to 0.5, i.e.,

$$\begin{cases} g_i^* = 0.5\hat{g}_i + 0.5, & i = 1,2,\cdots,50 \\ f_i^* = 0.5\hat{f}_i + 0.5, & i = 1,2,\cdots,50 \\ g_{0,v}^* = 0.5\hat{g}_{0,v} + 0.5, & v = 1,2,\cdots,40 \end{cases} \quad (30)$$

Then, LQD transformation was applied to PDFs of Eq. (30) from Eq. (8) to obtain $\{\psi_i^{g^*}\}_{i=1}^{50}$, $\{\psi_i^{f^*}\}_{i=1}^{50}$ and $\{\psi_v^{g_0^*}\}_{v=1}^{40}$ as shown in Fig. 6 (a), (b) and (c), respectively.

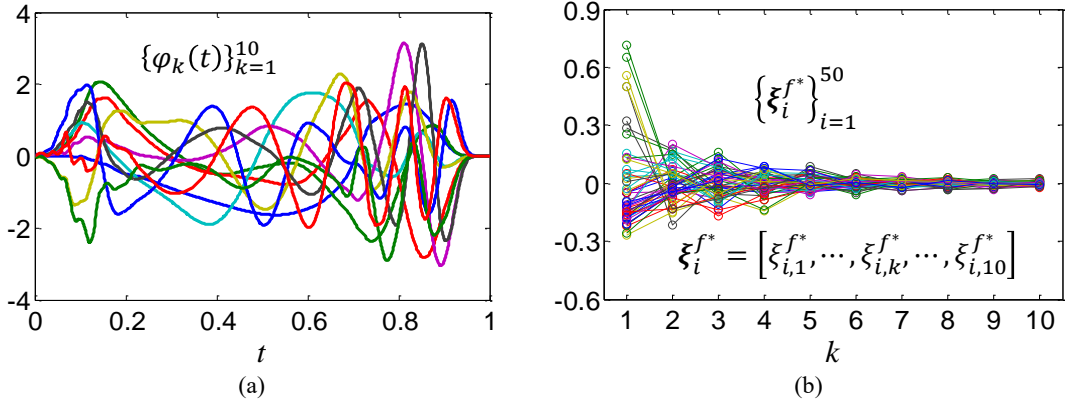

**Fig. 7.** FPCA on the functional dataset $\{\psi_i^{f^*}\}_{i=1}^{50}$: (a) eigenfunctions corresponding to the 10 largest eigenvalues and (b) representation vectors obtained via FPCA-based dimension reduction.

The FPCA was applied to $\{\psi_i^{f^*}\}_{i=1}^{50}$ for dimension reduction. Continuous functional data $\{\psi_i^{f^*}\}_{i=1}^{50}$ can be approximated by the truncated Karhunen–Loève representation using Eq. (15), and results show that 10-truncated Karhunen–Loève representation can effectively approximate such functions. Therefore, the truncation order $m$ was set as 10, and the 10 eigenfunctions are shown in Fig. 7 (a). Representation vectors ($\boldsymbol{\xi}_i^{f^*} = [\xi_{i,1}^{f^*},\cdots,\xi_{i,k}^{f^*},\cdots,\xi_{i,10}^{f^*}]$, $i = 1,\cdots,50$) of the functional data $\{\psi_i^{f^*}\}_{i=1}^{50}$ are shown in Fig. 7 (b).

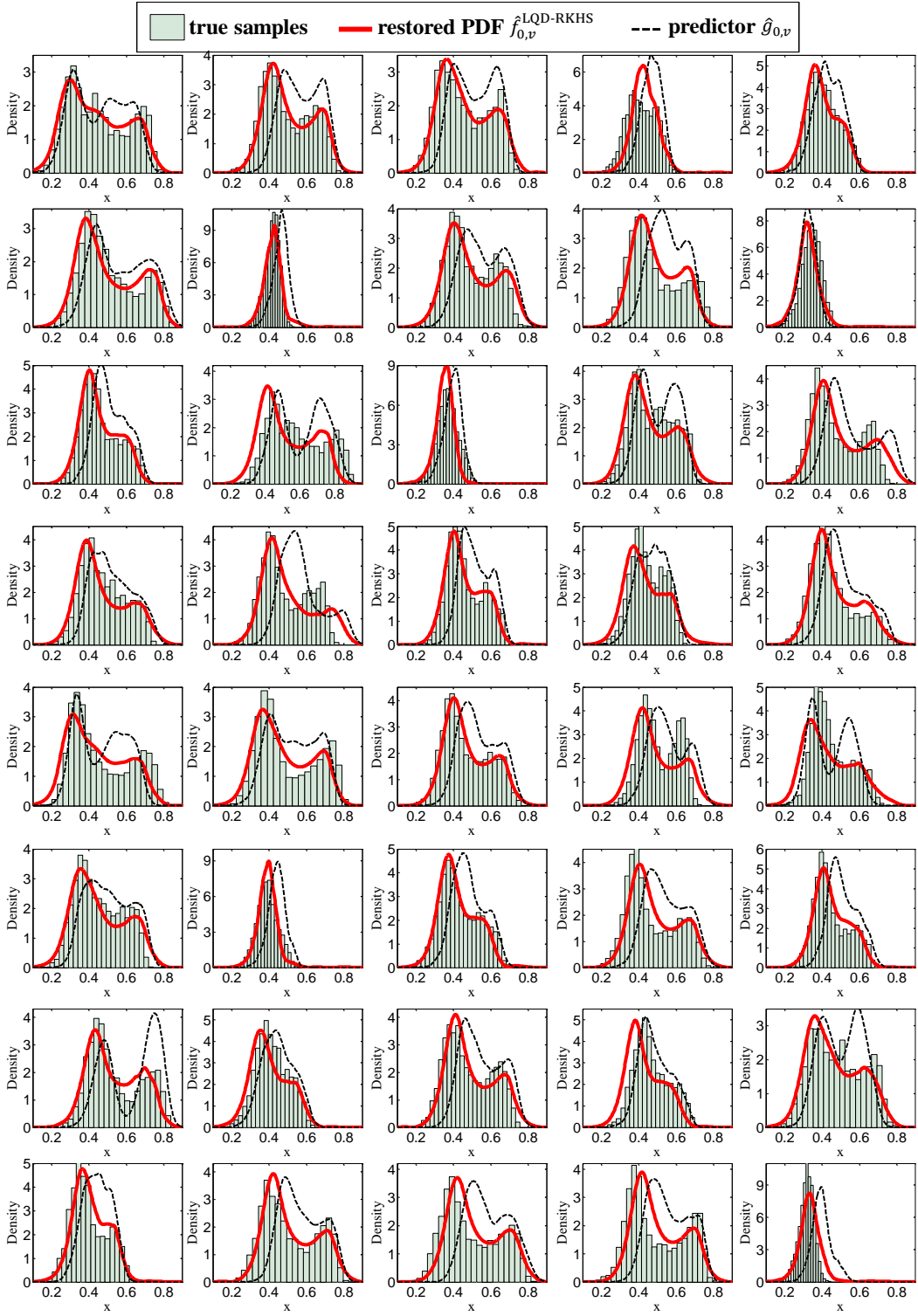

**Fig. 8.** Results of PDFs restored for the 40 test PDFs.

The RKHS-based function-to-vector regression model was used to restore the representation vectors $\left\{\boldsymbol{\xi}_v^{f_0^*}\right\}_{v=1}^{40}$. In the regression model, the Gaussian operator kernel of Eq. (20) was used. Parameter $\sigma$ of the Gaussian operator kernel was calculated by training data as shown in [31]

$$\sigma = \frac{1}{50\times 50}\sum_{i=1}^{50}\sum_{j=1}^{50}\sqrt{d_{ij}}, \text{ where } d_{ij} = \int \left|\psi_i^{g^*}(\tau) - \psi_j^{g^*}(\tau)\right|^2 d\tau \tag{31}$$

Additionally, smoothing parameter $\lambda_s$ of the RKHS-based regression model was set as 0.1 (it can also be selected through cross-validation as shown in Appendix 3).

After the representation vectors $\{\xi_v^{f_0^*}\}_{v=1}^{40}$ were restored by the regression model, the missing PDFs were further restored using the procedure discussed in section 4.5, and the restored PDFs of the 40 test PDFs are denoted as $\{\hat{f}_{0,v}^{\text{LQD-RKHS}}\}_{v=1}^{40}$. For comparison, the 40 restored PDFs are shown in Fig. 8, where the true missing samples, restored PDFs and PDFs of the collaborative sensor are denoted by histograms, bold curves and dashed curves, respectively. It can be seen that most of the restored PDFs agree well with the histograms of true missing samples. The effectiveness of the proposed LQD-RKHS method is verified.

**5.3 Performance evaluation for general cases**

To evaluate the performance of the LQD-RKHS method, the conventional distribution-to-distribution (DDR) regression method [25] and the distribution-to-warping function regression (DWR) [8] method were applied for purposes of comparison. A brief introduction of the DDR and DWR methods is given in Appendix 2. Fifty repeated tests were conducted to draw performance comparisons with each test involving 50 pairs of randomly selected training PDFs and 100 randomly selected test PDFs.

The mean integrated absolute error (MIAE) is used to quantify the error of restored PDFs through a single test involving 100 test PDFs. Let $\{\hat{f}_{0,v}\}_{v=1}^{100}$ and $\{\hat{f}_{0,v}^{\text{LQD-RKHS}}\}_{v=1}^{100}$ be the 100 randomly selected test PDFs and corresponding restored PDFs obtained from the LQD-RKHS method, respectively. The integrated absolute error (IAE) of a given test PDF $\hat{f}_{0,v}$ is defined as

$$\text{IAE}(v) = \int |\hat{f}_{0,v}^{\text{LQD-RKHS}}(\tau) - \hat{f}_{0,v}(\tau)|d\tau \tag{32}$$

For the 100 randomly selected test PDFs $\{\hat{f}_{0,v}\}_{v=1}^{100}$, the mean integrated absolute error (MIAE) of the LQD-RKHS method is calculated as

$$\text{MIAE(LQD-RKHS)} = \frac{1}{100}\sum_{v=1}^{100}\int |\hat{f}_{0,v}^{\text{LQD-RKHS}}(\tau) - \hat{f}_{0,v}(\tau)|d\tau \tag{33}$$

MIAEs of the DWR and DDR methods can be calculated similarly and are denoted as MIAE(DWR) and MIAE(DDR), respectively. For performance comparisons, relative MIAEs of the DWR and LQD-RKHS methods with respect to the DDR method are further defined as

$$R_{\text{MIAE}}(\text{DWR}) = \frac{\text{MIAE(DWR)}}{\text{MIAE(DDR)}} \ , \ R_{\text{MIAE}}(\text{LQD-RKHS}) = \frac{\text{MIAE(LQD-RKHS)}}{\text{MIAE(DDR)}} \tag{34}$$

Both the DDR and DWR methods are classical kernel regression-based methods with corresponding regression functions estimated from the Nadaraya-Watson kernel estimator. For the classical kernel regression, the $L_1$ distance (i.e., $\delta(\hat{g}_0, \hat{g}_i) = \int |\hat{g}_0(\tau) - \hat{g}_i(\tau)| d\tau$) is used as the similarity measure for both methods. Two kernel functions (the Gaussian kernel $K(u) = \exp(-u^2/2)/\sqrt{2\pi}$ and triangular kernel $K(u) = (1 - |u|)I_{\{|u| \leq 1\}}$) are considered in the kernel regression. Generally, two regression strategies are used for a classical kernel regression (involving the DDR and DWR methods): kernel regression with automatic bandwidth selection where bandwidth $h$ is directly selected from the cross-validation procedure presented in Appendix 3 and kernel regression involving the automatic selection of the number of neighbours where bandwidth $h$ is indirectly determined by the following equation (Eq. (35)) with parameter $\zeta$ directly selected from the cross-validation procedure presented in Appendix 3

$$h_0 = \min\left\{ h > 0 \,\middle|\, \sum_{i=1}^{n} I\left\{ K\left( \frac{\delta(\hat{g}_0, \hat{g}_i)}{h} \right) > 0 \right\} \geq n \times (\zeta\%) \right\} \quad (35)$$

where $n$ is the number of training distribution pairs, $I\{\cdot\}$ is the indicator function, $\delta(\cdot,\cdot)$ is the similarity measure, $K(\cdot)$ is the kernel function with finite support near the origins of coordinates (e.g., the triangular kernel). The number of neighbours of predictor $\hat{g}_0$ is represented by $n \times (\zeta\%)$. For a more detailed discussion, please refer to [8].

The cross-validation procedure presented in Appendix 3 was also applied to select the kernel function and regression strategy used for the DDR and DWR methods. Based on their performance in the cross-validation of the 30 repeated tests (each test involved 50 pairs of randomly selected training PDFs), the DDR method adopted the Gaussian kernel using the regression strategy of automatically selected bandwidth while the DWR method adopted the triangular kernel with the regression strategy involving automatically selecting the number of neighbours. The above optimization measures were applied to unlock the potential of the competitive methods.

For the LQD-RHKS method, the Gaussian operator kernel defined in Eq. (20) was selected with parameter $\sigma$ determined by Eq. (31), and $\lambda_s$ was fixed at 0.1 (the same parameter value used for the effectiveness validation test described in section 5.2).

The calculated relative MIAEs (by Eq. (34)) of the 50 repeated tests (each test involved 50 pairs of randomly selected training PDFs and 100 randomly selected test PDFs) are shown in Fig. 9 (a) and corresponding boxplots are shown in Fig. 9 (b). It can be observed from Fig. 9 (a) that the relative MIAEs of the DWR method are valued at less than 1 for 37 of the 50 tests. Therefore, DWR is superior to DDR by 37-13. All relative MIAEs of the LQD-RKHS method are valued at less than 1, and only one test (No. 35) involving the LQD-RKHS method produced worse results than DWR. Thus, the LQD-RKHS method is better than DDR and DWR by 50-0 and 49-1, respectively. The boxplots also show that the proposed LQD-RKHS method performs much better than the two competitive methods.

On the other hand, as noted in the introduction, the DWR method is relatively inefficient because continuous warping functions must be obtained from a function space using the optimization algorithm. However, when applying the LQD-RKHS method, analytic solutions are available through LQD transformation, FPCA-based dimension reduction and RKHS-based function-to-vector regression. Therefore, the LQD-RKHS method is also more efficient than the DWR method.

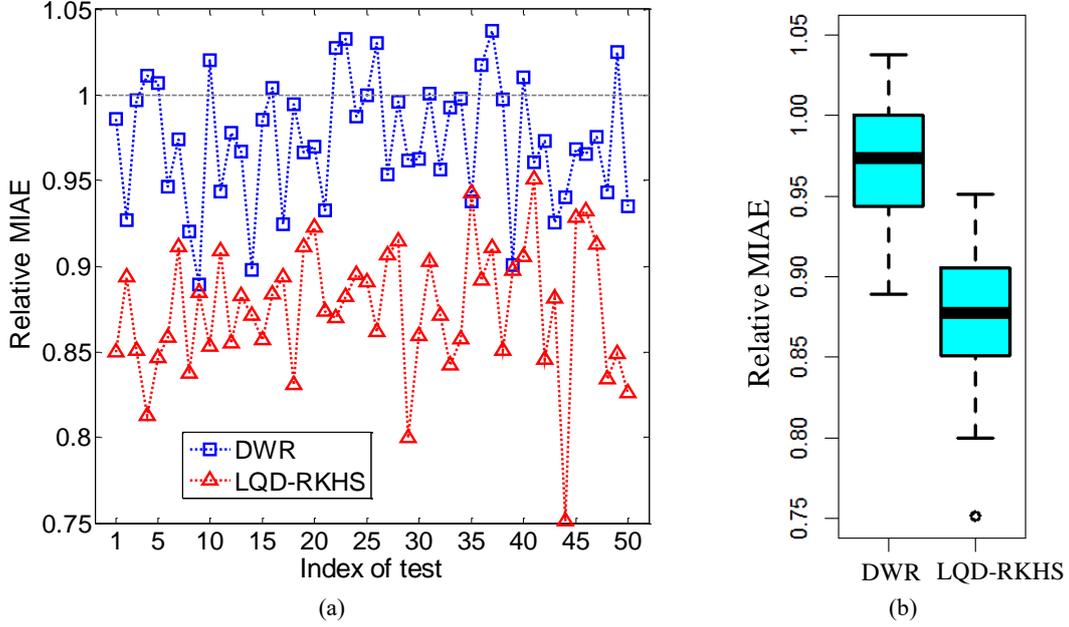

**Fig. 9.** Quantitative (error) comparisons of the DWR and LQD-RKHS methods for the 50 repeated tests designed for performance evaluation: (a) calculated relative MIAEs of the 50 repeated tests for each method and (b) box plot of the 50 calculated MIAEs for each method.

### 5.4 Performance evaluation for extrapolation cases

In this section, a test was conducted to investigate the performance of the LQD-RKHS method in terms of extrapolation predictions and to compare its performance to the DWR method. The DDR method was not considered here because DDR is limited in terms of extrapolation. The same training and test PDFs as those used in [8] and shown in Fig. 10 were applied.

For the LQD-RHKS method, the Gaussian operator kernel with parameters determined by Eq. (31) was also used, and smoothing parameter $\lambda_s$ was set as 0.15 (determined by cross-validation). For the six test PDFs, regression results obtained from the LQD-RKHS and DWR given in [8] are shown in Fig. 11. Fig. 11 shows that the LQD-RKHS method performed worse than the DWR method, showing that the LQD-RKHS method is limited in terms of extrapolation.

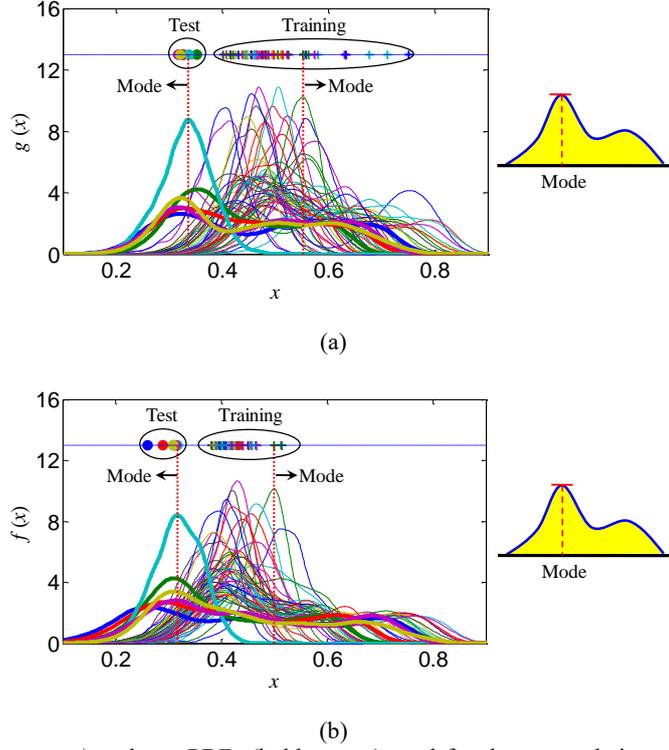

(a)

(b)

**Fig. 10.** Training (light curves) and test PDFs (bold curves) used for the extrapolation test [8] (the mode of a continuous PDF is the value x at which the PDF reaches its maximum): (a) Strain gauge A and (b) Strain gauge B.

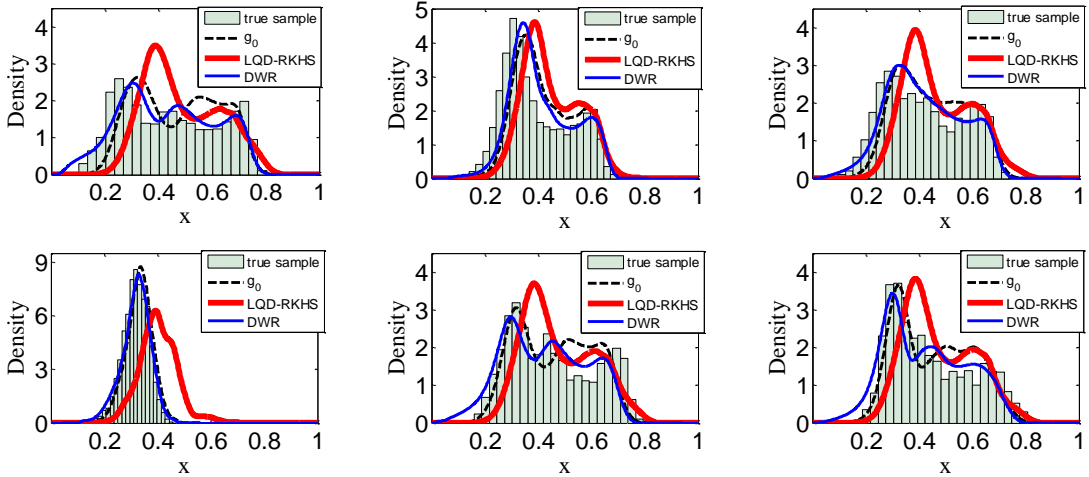

**Fig. 11.** Comparisons of results obtained from the LQD-RKHS and DWR [8] via extrapolation.

## Conclusions

This article proposes a new indirect distribution-to-distribution regression approach, i.e., the log-quantile-density (LQD) Reproducing Kernel Hilbert Space (RKHS) method, for restoring distributions of incomplete samples employed for structural health monitoring (SHM). The approach takes advantage of LQD transformation and nonparametric regression tools of RKHS to improve the precision and efficiency of missing distribution restoration. The effectiveness and performance of the method are investigated from filed monitoring data. The following conclusions

are drawn:

(1) The integral of the inverse LQD transformation calculated numerically may introduce considerable errors for certain probability density functions (PDFs) with values of close to zero within a certain sub-interval; preconditioning is proposed as a means to address this problem, which involves adding a uniform PDF.

(2) The dimension reduction of representation functions of distributions corresponding to an intermittent sensor improves the scalability of the RKHS-based functional regression model. The results of experiments described in this article show that 10-truncated Karhunen–Loève representation is sufficient to capture the main features of the continuous functions investigated.

(3) When the training dataset used is not too large, the proposed LQD-RKHS method is highly efficient and benefits from analytical solutions of the mathematical methodologies involved; however, for massive training data, solving the RKHS-based regression model from all training data presents challenges of matrix inversion, in this case, a local fitting strategy is preferred.

(4) Test results derived from strain monitoring data for an in-service bridge show that the proposed LQD-RKHS method performs much better than the conventional distribution-to-distribution regression (DDR) and distribution-to-warping function regression (DWR) methods in most cases; in cases involving extrapolation, the DWR method performs better.

**Acknowledgements**

This work was financially supported by the National Natural Science Foundation of China (Grant Nos. 51638007, U1711265 and 51678203), the Ministry of Science and Technology of China (Grant No. 2015DFG82080), and the China Scholarship Council (CSC).

**References**

[1] J. Ou, H. Li, Structural health monitoring in mainland China: review and future trends, Struct. Health Monit. 9(3) (2010) 219-231.

[2] F. Magalhães, A. Cunha, E. Caetano, Vibration based structural health monitoring of an arch bridge: from automated OMA to damage detection, Mech. Syst. Signal Process, 28 (2012) 212-228.

[3] E.J. Cross, K.Y. Koo, J.M.W. Brownjohn, K. Worden, Long-term monitoring and data analysis of the Tamar Bridge, Mech. Syst. Signal Process 35(1) (2013) 16-34.

[4] V.G.M. Annamdas, S. Bhalla, C.K. Soh, Applications of structural health monitoring technology in Asia, Struct. Health Monit. 16(3) (2017) 324-346.

[5] Y. Yang, S. Nagarajaiah, Harnessing data structure for recovery of randomly missing structural vibration responses time history: Sparse representation versus low-rank structure, Mech. Syst. Signal Process 74 (2016) 165-182.

[6] Z. Zhang, Y. Luo, Restoring method for missing data of spatial structural stress monitoring

based on correlation, Mech. Syst. Signal Process 91 (2017) 266-277.

[7] Z. Chen, Y. Bao, H. Li, Imputing missing strain monitoring data in structural health monitoring, in: Proceedings of the 11th international workshop on structural health monitoring, Stanford, CA, 12-14, September 2017, pp. 993-999.

[8] Z. Chen, Y. Bao, H. Li, B.F. Spencer Jr., A novel distribution regression approach for data loss compensation in structural health monitoring, Struct. Health Monit. (2017) doi: 10.1177/1475921717745719.

[9] J.A. Goulet, K. Koo, Empirical validation of Bayesian dynamic linear models in the context of structural health monitoring, J. Bridge Eng. 23(2) (2017) 05017017.

[10] Y. Bao, Z. Tang, H. Li, Y. Zhang, Computer vision and deep learning–based data anomaly detection method for structural health monitoring, Struct. Health Monit. (2018) doi: 10.1177/1475921718757405.

[11] T. Nagayama, S.H. Sim, Y. Miyamori, B.F. Spencer, Issues in structural health monitoring employing smart sensors, Smart Struct. Syst. 3(3) (2007) 299–320.

[12] S. Wei, Z. Zhang, S. Li, H. Li, Strain features and condition assessment of orthotropic steel deck cable-supported bridges subjected to vehicle loads by using dense FBG strain sensors, Smart Mater. Struct. 26(10) (2017) 104007.

[13] S. Li, S. Wei, Y. Bao, H. Li, Condition assessment of cables by pattern recognition of vehicle-induced cable tension ratio, Eng. Struct. 155 (2018) 1-15.

[14] H. Xia, Y. Ni, K. Wong, J.M. Ko, Reliability-based condition assessment of in-service bridges using mixture distribution models, Comput. Struct. 106-107 (2012) 204-213.

[15] Z. Chen, Y. Xu, Y. Xia, Q. Li, K. Wong, Fatigue analysis of long-span suspension bridges under multiple loading: Case study, Eng. Struct. 33(12) (2011) 3246-3256.

[16] N. Lu, M. Noori, Y. Liu, Fatigue reliability assessment of welded steel bridge decks under stochastic truck loads via machine learning, J. Bridge Eng. 22(1) (2017) 04016105.

[17] N.M. Okasha, D.M. Frangopol, Integration of structural health monitoring in a system performance based life-cycle bridge management framework, Struct. Infrastruct. E. 8(11) (2012) 999-1016.

[18] E.J. OBrien, F. Schmidt, D. Hajializadeh, X.Y. Zhou, B. Enright, C.C. Caprani, S. Wilson, E. Sheils, A review of probabilistic methods of assessment of load effects in bridges, Struct. Saf. 53 (2015) 44-56.

[19] Y. Bao, H. Li, X. Sun, Y. Yu, J. Ou, Compressive sampling-based data loss recovery for wireless sensor networks used in civil structural health monitoring, Struct. Health Monit. 12(1) (2013) 78-95.

[20] Y. Bao, Y. Yu, H. Li, X. Mao, W. Jiao, Z. Zou, J. Ou, Compressive sensing-based lost data recovery of fast-moving wireless sensing for structural health monitoring, Struct. Control. Health Monit. 22(3) (2015) 433-448.

[21] Z. Zou, Y. Bao, H. Li, B.F. Spencer Jr., J. Ou, Embedding compressive sensing-based data loss recovery algorithm into wireless smart sensors for structural health monitoring, IEEE Sens. J. 15(2) (2015) 797-808.

[22] X. Zhao, J. Jia, Y. Zheng, Strain monitoring data restoring of large-span steel skybridge based on BP neural network, Chin. J. Archit. Civil Eng. 26(1) (2009)101-106.

[23] Y. Huang, D. Wu, J. Li, Structural healthy monitoring data recovery based on extreme learning machine, Chin. Comput. Eng. 37(16) (2011) 241-243.

[24] Y. Huang, D. Wu, Z. Liu, J. Li, Lost strain data reconstruction based on least squares support vector machine, Chin. Meas Control Tech 29 (2010) 8-12.

[25] J.B. Oliva, B. Póczos, J. Schneider, Distribution to distribution regression, in: Proceedings of the 30th International Conference on Machine Learning, Atlanta, GA, 16-21, June 2013, pp. 1049-1057.

[26] A. Petersen, H.G. Müller, Functional data analysis for density functions by transformation to a Hilbert space, Ann. Statist. 44(1) (2016) 183-218.

[27] C.H. Lampert, Predicting the future behavior of a time-varying probability distribution, in Proceedings of the IEEE conference on computer vision and pattern recognition, 2015, pp. 942-950.

[28] F.M.L. Di Lascio, S. Giannerini, A. Reale, Exploring copulas for the imputation of complex dependent data, Stat. Method Appl. 24(1) (2015) 159–175.

[29] S. Dasgupta, D. Pati, A. Srivastava, A geometric framework for density modeling, https://arxiv.org/abs/1701.05656v1, 2017.

[30] B.C. Turnbull, S.K. Ghosh, Unimodal density estimation using Bernstein polynomials, Comput. Statist. Data Anal. 72 (2014) 13-29.

[31] H. Lian, Nonlinear functional models for functional responses in reproducing kernel Hilbert spaces, Canad. J. Statist. 35(4) (2007) 597-606.

[32] J.O. Ramsay, B.W. Silverman, Functional data analysis (Second Edition), Springer, New York, 2005.

[33] M. Benko, W. Härdle, A. Kneip, Common functional principal components, Ann. Statist.37(1) (2009) 1-34.

[34] J.L. Bali, G. Boente, D.E. Tyler, J.L. Wang, Robust functional principal components: A projection-pursuit approach, Ann. Statist. 39(6) (2011) 2852-2882.

[35] T.T. Cai, M. Yuan, Minimax and adaptive prediction for functional linear regression, J. Amer. Statist. Assoc. 107(499) (2012) 1201-1216.

[36] H. Kadri, E. Duflos, P. Preux, S. Canu, A. Rakotomamonjy, J. Audiffren, Operator-valued kernels for learning from functional response data, J. Mach. Learn. Res. 17 (2016) 1-54.

[37] B. Schölkopf, R. Herbrich, A. Smola, A generalized representer theorem, in Computational learning theory, Springer: Berlin/Heidelberg, 2001, pp. 416-426.

[38] B.W. Silverman, Density estimation, Chapman and Hall, London,1986.

[39] F. Ferraty, P. Vieu, Nonparametric functional data analysis: theory and practice, Springer, New York, 2006.

**Appendix 1: Solving FPCA problems by discretization**

The discretization method for solving FPCA is described in Ramsay and Silverman's book (Section 8.4) [32].

Consider the centred functional dataset presented in Eq. (9), i.e., $\left\{\tilde{\psi}_i^{f^*}(t)\right\}_{i=1}^n$, $t \in [0, 1]$. Let $\{t_l\}_{l=1}^{T_p}$ be an evenly spaced fine partition of the time interval $[0, 1]$ such that $0 = t_1 < t_2 < \cdots < t_{T_p} = 1$. Then, a continuous function $\tilde{\psi}_i^{f^*}(t)$ can be roughly represented by a vector, i.e., $\boldsymbol{x}_i = \left[\tilde{\psi}_i^{f^*}(t_1) \ \tilde{\psi}_i^{f^*}(t_2) \ \cdots \ \tilde{\psi}_i^{f^*}\left(t_{T_p}\right)\right]^\mathrm{T} \in \mathrm{R}^{T_p \times 1}$. The investigated centred functional dataset, i.e., $\left\{\tilde{\psi}_i^{f^*}(t)\right\}_{i=1}^n$, can be transformed into the following matrix

$$\mathbf{X} = [\boldsymbol{x}_1 \ \boldsymbol{x}_2 \ \cdots \ \boldsymbol{x}_n]^T \in \mathrm{R}^{n \times T_p} \tag{36}$$

The sample covariance matrix of $\mathbf{X}$ is $\mathbf{V} = n^{-1}\mathbf{X}^T\mathbf{X}$, and it can be verified that the element of matrix $\mathbf{V}$ is the value of the covariance function defined in Eq. (10) at grid points, i.e.,

$$\mathbf{V} = \{V_{jl}\}_{j=1 \ l=1}^{T_p \ \ T_p} = \{v(t_j, t_l)\}_{j=1 \ l=1}^{T_p \ \ T_p} \tag{37}$$

An ordinary principal component analysis (PCA) of $n \times T_p$ data matrix $\mathbf{X}$ can be realized by solving the following eigenanalysis problem

$$\mathbf{V}\boldsymbol{u} = \lambda \boldsymbol{u} \text{ subject to } \boldsymbol{u}^T\boldsymbol{u} = 1 \tag{38}$$

where $\lambda$ and $\boldsymbol{u}$ are the eigenvalue and eigenvector, respectively.

After an ordinary PCA has been carried out for the discrete data, the result of the ordinary PCA problem (see Eq. (38)) is transformed into the corresponding result of the FPCA problem (see Eq. (12)). Let $w$ be the length of the evenly spaced subinterval of the discrete time, i.e., $w = t_l - t_{l-1}$ ($l = 2,3,\cdots,T_p$). Recall the covariance operator defined in Eq. (11) for time point $t_j$ derived from the principle of numerical integration:

$$V_C\varphi(t_j) = \int v(t_j, s)\varphi(s)ds \approx w \sum_{l=1}^{T_p} v(t_j, t_l)\varphi(t_l) \tag{39}$$

Then, the functional eigenanalysis problem presented in Eq. (12), i.e., $V_C\phi = \rho\phi$, can be approximated as the following discrete form

$$w\mathbf{V}\boldsymbol{\varphi}_d = \rho\boldsymbol{\varphi}_d \tag{40}$$

where vector $\boldsymbol{\varphi}_d$ is formed by values of eigenfunction $\varphi(t)$ at discrete time points, i.e., $\boldsymbol{\varphi}_d = [\varphi(t_1) \ \varphi(t_2) \ \cdots \ \varphi(t_{T_p})]^T$. The normalization constraint of eigenfunction $\varphi(t)$ can also approximated by a discrete form as

$$\|\varphi\|^2 = \int \varphi^2(t)dt \approx w \sum_{l=1}^{T_p} \varphi^2(t_l) = w(\boldsymbol{\varphi}_d)^T\boldsymbol{\varphi}_d = 1 \tag{41}$$

By comparing Eqs. (38), (40) and (41), it has $\rho = w\lambda$ and $\boldsymbol{\varphi}_d = w^{-1/2}\boldsymbol{u}$ where $\boldsymbol{u}$ is the normalized eigenvector of the normal multivariate PCA (see Eq. (38)). Note that because $\boldsymbol{\varphi}_d = [\varphi(t_1) \ \varphi(t_2) \ \cdots \ \varphi(t_{T_p})]^T$, the continuous eigenfunction $\varphi(t)$ of the FPCA problem can be obtained by interpolation using corresponding nodal values represented by vector $\boldsymbol{\varphi}_d$.

## Appendix 2: DDR and DWR methods

For training PDFs $\{\hat{g}_i, \hat{f}_i\}_{i=1}^n$ and predictor $\hat{g}_0$, the restoration of missing PDF $f_0$ obtained from DDR is

$$\hat{f}_0^{\text{DDR}} = F_{reg}^{\text{DDR}}(\hat{g}_0) = \sum_{i=1}^n \frac{K(\delta(\hat{g}_0, \hat{g}_i)/h)}{\sum_{j=1}^n K(\delta(\hat{g}_0, \hat{g}_j)/h)} \hat{f}_i \tag{42}$$

where $K(\cdot)$ is the kernel function satisfying $K(u) \geq 0, \forall u \in \mathrm{R}$, $h > 0$ is the bandwidth, and $\delta(\hat{g}_0, \hat{g}_i)$ is a similarity measure of distributions. For a more detailed discussion about this distribution regression model, please refer to [25].

Suppose that the investigated PDFs are one-dimensional continuous PDFs finitely supported on $[0, 1]$. For training PDFs $\{\hat{g}_i, \hat{f}_i\}_{i=1}^n$, let $\{\hat{\gamma}_i\}_{i=1}^n$ be estimated warping functions used to transform $\hat{g}_i$ to approximately reach $\hat{f}_i$, i.e., $\hat{f}_i(x) \approx \hat{g}_i(\hat{\gamma}_i(x))\hat{\gamma}_i'(x), i = 1,2,\cdots,n$. Then, the restoration of missing PDF $f_0$ is obtained from $\hat{f}_0^{\text{DWR}}(x) = \hat{g}_0\left(\hat{\gamma}_0^{\text{DWR}}(x)\right)\hat{\gamma}_0^{\text{DWR}}{}'(x)$ via DWR where warping function $\hat{\gamma}_0^{\text{DWR}}$ is estimated from the kernel distribution-to-warping function regression model using $\hat{g}_0$ as the predictor, i.e.,

$$\hat{\gamma}_0^{\text{DWR}} = F_{reg}^{\text{DWR}}(\hat{g}_0) = \sum_{i=1}^n \frac{K(\delta(\hat{g}_0, \hat{g}_i)/h)}{\sum_{j=1}^n K(\delta(\hat{g}_0, \hat{g}_j)/h)} \hat{\gamma}_i \quad (43)$$

For a more detailed discussion of this distribution regression model, please refer to [8].

## Appendix 3: Hyperparameter selection or model optimization in nonparametric regression using cross-validation

Cross-validation is a trial approach to hyperparameter selection or model optimization using training data. Consider a distribution regression model with hyperparameter $\theta$ (e.g., bandwidth $h$ of the classical kernel regression, smoothing parameter $\lambda_s$ of RKHS-based regression, etc.) and training PDFs $\{\hat{g}_i, \hat{f}_i\}_{i=1}^n$. For the cross-validation procedure, a prediction of PDF $\hat{f}_k(x), k \in \{1,2,\cdots,n\}$ of the set of training PDFs is obtained from the leave-one-out nonparametric regression model for distributions, i.e.,

$$\{\hat{f}_i(x), \hat{g}_i(x)\}_{i=1,i\neq k}^n \cup \hat{g}_k(x) \xrightarrow{distribution\ regression} \hat{f}_k^{reg}(x|\theta),\ k = 1,\cdots,n \quad (44)$$

where $\hat{f}_k^{reg}$ is the prediction of $\hat{f}_k$ obtained from the distribution regression model using $\{\hat{f}_i(x), \hat{g}_i(x)\}_{i=1,i\neq k}^n$ as training distributions and $\hat{g}_k$ as the predictor. Then, hyperparameter $\theta$ can be automatically selected from the following risk minimization principle

$$\hat{\theta}_{opt} = \underset{\theta}{\text{argmin}} \sum_{k=1}^n \int \left(\hat{f}_k^{reg}(\tau|\theta) - \hat{f}_k(\tau)\right)^2 d\tau \quad (45)$$

This cross-validation procedure can also be applied to select appropriate kernel functions or other specific regression strategies for model optimization. For a more detailed discussion of cross-validation in nonparametric regression, please refer to [39].